\documentclass[conference]{IEEEtran}
\IEEEoverridecommandlockouts
\usepackage[utf8]{inputenc}
\usepackage[T1]{fontenc}
\usepackage{amsmath,amssymb,amsfonts,times}
\usepackage{graphicx}
\usepackage{textcomp}
\usepackage{pgfplots}
\usepackage{pdfpages}
\usepackage{scalefnt}
\newcommand{\norm}[1]{\left\lVert#1\right\rVert}

\graphicspath{{figures/}}
\def\BibTeX{{\rm B\kern-.05em{\sc i\kern-.025em b}\kern-.08em
    T\kern-.1667em\lower.7ex\hbox{E}\kern-.125emX}}
\begin{document}

\title{Study of Switched Max-Link Buffer-Aided Relay Selection for Cooperative MIMO Systems \\
}

\author{\IEEEauthorblockN{F. L. Duarte$^{1, 2}$ and R. C. de Lamare$^{1, 3}$}
\IEEEauthorblockA{$^{1}$Centre for Telecommunications Studies (CETUC), Pontifical Catholic University of Rio de Janeiro, Brazil \\
$^{2}$Military Institute of Engineering, IME, Rio de Janeiro, RJ, Brazil  \\
$^{3}$Department of Eletronic Engineering, University of York, United Kingdon \\
Email: \{flaviold, delamare\}@cetuc.puc-rio.br}
}

\maketitle
\begin{abstract}
In this paper, we investigate relay selection for cooperative
multiple-antenna systems that are equipped with buffers, which
increase the reliability of wireless links. In particular, we
present a novel relay selection technique based on switching and the
Max-Link protocol that is named Switched Max-Link. We also introduce
a novel relay selection criterion based on the maximum likelihood
(ML) principle denoted maximum minimum distance that is incorporated
into. Simulations are then employed to evaluate the performance of
the proposed and existing techniques.  \\
\end{abstract}


\section{INTRODUCTION}
In wireless networks, signal fading caused by multipath propagation
is a channel damage that can be mitigated through the use of
diversity \cite{f1,f2,f3}. Spatial diversity techniques are
attractive as they can be combined with other forms of diversity. In
cooperative communications with multiple relays, where a number of
relays help a source in transmitting data packets to a destination,
by receiving, processing (decoding) and forwarding these packets,
relay selection schemes are key because of their high performance
\cite{f5,f4,armo,smce}.

As cooperative communication can improve the throughput and extend
the coverage of wireless communications systems, the task of relay
selection serves as a building block to realize it. Simple relay
schemes have been/are being included in recent/future wireless
standards such as Long Term Evolution (LTE) Advanced \cite{f6} and
5G standards. In conventional relaying, using half duplex (HD) and
decode-and-forward protocols, transmission is usually organized in a
prefixed schedule with two successive time slots. In the first time
slot, the relay receives and decode the data transmitted from the
source, and in the second time slot the relay forwards the decoded
data to the destination.  Single relay selection schemes use the
same relay for reception and transmission, so they are not able to
simultaneously exploit the best available source-relay (SR) and
relay-destination (RD) channels. The two most common schemes are
bottleneck based and maximum harmonic mean based best relay
selection (BRS).

Since the relays use a prefixed schedule of transmission and
reception independent of the time-varying quality of the channels in
wireless systems, this may lead to performance degradation
\cite{f6}. Because of the fixed schedule of transmission and
reception for the relays, the system cannot exploit the best
source-relay (SR) and the best relay-destination (RD) channels (the
links with the highest powers). The performance could be improved if
the link with the highest power could be used in each time slot.
This can be achieved via a buffer-aided relaying protocol that does
not have a prefixed schedule of reception and transmission for the
relay. As the relay has to accumulate packets in its buffer, before
transmitting, selecting the link with the largest power may not
always be possible. The use of buffers provides an improved
performance and new degrees of freedom for system design. However,
it faces a challenge: the introduction of additional delay that must
be well managed for delay-sensitive applications. Buffer-aided
relaying protocols \cite{f6,f7,f9,badstc,ba_plnc} require not only
the acquisition of channel state information (CSI), but control of
the buffer status \cite{f6}. Some possible applications of
buffer-aided relaying are: vehicular, cellular, and sensor networks
\cite{f6}.

In Max-Max Relay Selection (MMRS) \cite{f5}, in the first time slot,
the relay selected for reception can store the received packets in
its buffer and forward them at a later time when selected for
transmission. In the second time slot, the relay selected for
transmission can transmit the first packet in the queue of its
buffer, which was received from the source earlier. MMRS assumes
that the buffer of the relay selected for reception is never full
and the buffer of the relay selected for transmission is never
empty. It is only possible for infinite size buffers. Considering
finite buffer sizes, the buffer of a relay becomes empty if the
channel conditions are such that it is selected repeatedly for
transmission (and not for reception) or full if it is selected
repeatedly for reception (and not for transmission). To overcome
this limitation, in  \cite{f5} a hybrid relay selection (HRS) scheme
was proposed, which is a combination of conventional BRS and MMRS.
For HRS, if the buffer of the relay selected for reception is full
or if the buffer of the relay selected for transmission is empty,
BRS is employed; otherwise, MMRS is used. Similar to MMRS, in HRS
the relays selected for reception and transmission in successive
time slots may be different.

Although MMRS and HRS  improve the throughput and/or SNR gain as
compared to BRS, their diversity gain is limited to N (the quantity
of relays). This can be improved by combining adaptive link
selection with MMRS, which results in a protocol referred to as
Max-Link  \cite{f9}. The main idea of Max-Link is to select in each
time slot the strongest link among all the available SR and RD links
(i.e., among $2N$ links) for transmission. So each time slot is
allowed to be allocated dynamically to the source or relay
transmission,  according to the instantaneous quality of the links
and the status of the buffers \cite{f7}. For i.i.d. links and no
delay constraints, Max-Link achieves a diversity gain of $2N$, which
is double the diversity gain of BRS and MMRS. These relay selection
protocols were developed primarily for transmission without delay
constraints and relay networks with i.i.d. links. In general, a
delay-constrained protocol can be obtained by limiting the size of
the buffers. On the other hand, these protocols require full CSI of
all links at the destination. Hence, the protocols have similar
complexities and applicability \cite{f6}.

In this work, we examine buffer-aided relay selection for
cooperative multiple-antenna systems \cite{mmimo,wence}. In
particular, we combine the concept of switching with the Max-Link
protocol for cooperative multiple-antenna systems, which results in
the proposed Switched MIMO/Max-Link relay selection technique. We
also introduce the maximum minimum distance criterion for selection
of multiple relays, which is based on the maximum likelihood (ML)
criterion. Simulations illustrate the performance of the proposed
and analyzed relay selection techniques.

This paper is structured as follows. Section II describes the system
model and the main assumptions made. Section III presents the
proposed Switched Max-Link Relay Selection. Section IV illustrates
and discusses the simulation results whereas Section V gives the
concluding remarks.

\section{System Description}

We consider a relay network with one source node, $S$, one
destination node, $ D$, and N half-duplex decode and forward (DF)
relays, $R_1$,..., $R_N$. Each relay is equipped with a buffer and
each node is equipped with a quantity of $M$ antennas, and the
transmission is organized in time slots  \cite{f5}. The considered
system is shown in Fig. \ref{fig:model}.

\begin{figure}[!h]
\centering
\includegraphics[width=1\columnwidth]{fig1.eps}
\caption{System Model} \label{fig:model}
\end{figure}

\subsection{Assumptions}

In cooperative transmissions two time slots are needed to transmit
data packets from the source to the destination, as in direct
transmission just one time slot will be needed to do the same, so
the energy transmitted in direct transmission (from the source to
the destination) is twice the energy transmitted in the cooperative
transmission (from the source to the relay selected for reception or
from the relay selected for transmission to the destination). For
this reason, we have:

- The energy transmitted from each antenna in cooperative
transmissions (from the source to the relay selected for reception)
is $E_s/M$;

- The energy transmitted from each antenna in cooperative
transmissions (from the relay selected for transmission to the
destination) is  $E_{r_j}/M = E_s/M$ ;

- The energy transmitted from each antenna in direct transmissions
(from the source to the  destination) is  $2 \times E_s/M$.

We consider that the channel coefficients are mutually independent
zero mean complex Gaussian random variables (Rayleigh fading).
Moreover, we assume that the transmission is organized in data
packets and the channels are constant for the duration of one packet
and vary independently from one packet to the next.

The information about the order of the data packets is contained in
the preamble of each packet, so the original order is restored at
the destination node. Other information such as signaling for
network coordination and pilot symbols for training and knowledge of
the channel (CSI) are also inserted in the preamble of the packet.
Furthermore, we assume that the relays do not communicate with each
other.

\subsection{System Description}

The received signal is organized in an $M \times 1$ vector
$\mathbf{y_{s,d}}[i]$, transmitted from the antennas of the source
and collected by the receive antennas of the destination as given by
\begin{eqnarray}
    \mathbf{y}_{s,d}[i]= \sqrt{2 E_s/M} \mathbf{H}_{s,d}\mathbf{x}[i]+\mathbf{n}_d[i],
    \label{eq:1}
\end{eqnarray}
\noindent where $E_s$ represents the total energy of the symbols
transmitted from the source per time slot, $\mathbf{x}$ represents
the vector formed by $M$ symbols of the different packets being sent
by each of the antennas of the source (a symbol of each packet).
$\mathbf{H}_{s,d}$  represents the $M \times  M$ matrix of $SD$
links and $\mathbf{n}_d$  denotes the zero mean additive white
complex Gaussian noise (AWGN) at the destination.

The received signal from the source to the selected relay is
organized in an $M \times 1$ vector $\mathbf{y}_{s,r_k }[i]$ given
by
\begin{eqnarray}
    \mathbf{y}_{s,r_k}[i]=\sqrt{E_s/M} \mathbf{H}_{s,r_k}\mathbf{x}[i]+\mathbf{n}_{r_k}[i],
    \label{eq:2}
\end{eqnarray}
\noindent where $E_s$ represents the total energy of the symbols
transmitted from the source per time slot, $\mathbf{x}$ represents
the vector formed by $M$ symbols of the different packets being sent
by each of the antennas of the source (a symbol of each packet),
$r_k$ refers to the selected relay for reception,
$\mathbf{H}_{s,r_k}$ is the $M \times  M$ matrix of $SR_k$ links and
$\mathbf{ n}_{r_k}$ represents the AWGN at the relay selected for
reception.

The signal transmitted from the selected relay and received at the
destination is structured in an $M \times 1$ vector
$\mathbf{y}_{r_j,d }[i]$ given by
    \begin{eqnarray}
    \mathbf{y}_{r_j,d}[i]=\sqrt{E_{r_j}/M}  \mathbf{H}_{r_j,d}\hat{\mathbf{x}}[i]+\mathbf{n}_d[i],
    \label{eq:3}
    \end{eqnarray}
\noindent where $ E_{r_j}$ represents the total energy of the
decoded symbols transmitted from the relay selected for transmission
$r_j$ per time slot, $\hat{\mathbf{x}}[i]$ is the vector formed by $M$
previously decoded symbols in the relay selected for reception and
stored in its buffer and now transmitted by the relay selected for
transmission $r_j$, $\mathbf{H}_{r_j,d}$ is the $M \times M$ matrix
of  $R_jD$ links and $\mathbf{ n}_{d}$ is the AWGN at the
destination receiver.

At the relays, we employ the maximum likelihood (ML) receiver:
    \begin{eqnarray}
    \hat{\mathbf{x}}[i]= \arg \min_{\mathbf{x'}[i]} (\norm{\mathbf{y}_{s,r_k}[i]- \sqrt{E_s/M} \mathbf{H}_{s,r_k}\mathbf{x'}[i]}^2),
    \label{eq:4}
    \end{eqnarray}
where $\mathbf{x'}[i]$ represents each possibility of the vector
formed by $M$ symbols. As an example, if $m = 2$, $\mathbf{x'}[i]$
may be [0 0]$^T$, [0 1]$^T$, [1 0]$^T$ or [1 1]$^T$.

At the destination, we also resort to the ML receiver which
depending on the transmission ($SD$ or $ R_jD$) yields
    \begin{eqnarray}
    \hat{\mathbf{x}}[i]= \arg \min_{\mathbf{x'}[i]} (\norm{\mathbf{y}_{s,d}[i]- \sqrt{2 E_s/M} \mathbf{H}_{s,d}\mathbf{x'}[i]}^2),
    \label{eq:5}
    \end{eqnarray}
    \begin{eqnarray}
    \hat{\mathbf{x}}[i]= \arg \min_{\mathbf{x'}[i]} (\norm{\mathbf{y}_{r_j,d}[i]- \sqrt{ E_s/M} \mathbf{H}_{r_j,d}\mathbf{x'}[i]}^2),
    \label{eq:6}
    \end{eqnarray}
The ML receiver of the DF relay looks for an estimate of the vector
of symbols transmitted by the source  $\hat{\mathbf{x}}$ , comparing
the quadratic norm between the output $\mathbf{y}_{s,r_k}$ and the
term $ \sqrt{E_s}\mathbf{H}_{s,r_k}$  multiplied by
$\mathbf{x'}[i]$, that represents each  of the $2^M$  possibilities
of transmitted symbols vector $\mathbf{x}$ (considering BPSK). We
compute the symbol vector which is the optimal solution for the ML
rule. The same reasoning is applied to the ML receiver at the
destination. Other detection techniques can also be employed
\cite{delamare_mber,rontogiannis,delamare_itic,stspadf,choi,stbcccm,FL11,delamarespl07,jidf,jio_mimo,tds,peng_twc,spa,spa2,jio_mimo,P.Li,jingjing,memd,did,bfidd,mbdf,bfidd,mserrr,mmimo,wence,shaowcl08}.

\section{Proposed Switched Max-Link Relay Selection Protocol}

In this section, we detail the proposed Switched Max-Link relay
selection protocol for cooperative multiple-antenna systems. The
proposed Switched Max-Link  scheme can be implemented by making use
of a network with one source node, $S$,  one destination node, $ D$,
$N$ half-duplex DF relays, $R_1$,...,$R_N$, and the same number of
$M$ antennas in each node ($Na_s$ = $Na_r$ = $Na_d$ = $M$), forming
a number of $ M \times  N$ source-relay (SR) channels (links) for
reception, $ M \times  N$ relay-destination (RD) links for
transmission and  $ M$ direct source-destination links, as
illustrated in Fig. 1 \cite{f8}.

This scheme selects the best relay for reception ($r_k$) or the best
relay for transmission ($r_j$) between $N$ relays (the best set of
$M$ $SR$ links among $N$ sets or the best set of  $M$ $RD$ links
among $N$ sets). The relay selection criterion is based on the ML
criterion and looks for the maximum minimum distance,  which
corresponds to choosing the relay that has the highest minimum
distance and requires calculating the distance between the $2^m$
possible vector of transmitted symbols:

In each time slot, the proposed Switched Max-Link Relay Selection
Protocol may operate in two possible modes ("Direct Transmission" or
"Max-Link"), so this scheme has three options:

a) work in "Direct Transmission" mode, by the source sending a
quantity of $M$ packets directly to the destination;

b) work in "Max-Link" mode, by the source sending a quantity of $M$
packets to the relay selected for reception and these packets are
stored in its buffer;

c) work in "Max-Link" mode, by the relay selected for transmission
forwarding  a quantity of $M$ packets from its buffer to the
destination node.

Table 1 shows the Switched Max-Link pseudo-code and the following
subsections explain how this protocol works.

\begin{table}[h]
\centering
 \caption{Switched Max-Link  Pseudo-Code}
 \label{table1}
\begin{tabular}{l}

\hline
\\

$\mathbf{D}_{min}=[ ];$ \\

$\mathbf{for}$  i=1:N \\

  ~~  $\mathbf{D}_{SR}=[ ];$ \\

 ~~    $\mathbf{for}$  l= 1:$2^M - 1$ \\

   ~~~~     $\mathbf{for}$  n = l+1:$2^M$ \\

       ~~~~     $\mathcal{D}_{SR_i}=  \norm{\sqrt{ E_s/M} \mathbf{H}_{s,r_i}\mathbf{x}_l - \sqrt{ E_s/M} \mathbf{H}_{s,r_i}\mathbf{x}_n}^2$ ; \\

           ~~~~ $\mathbf{D}_{SR}=[ \mathbf{D}_{ SR}  \mathcal{D}_{SR_i}];$ \\

        ~~~~$\mathbf{end}$ \\

  ~~  $\mathbf{end}$ \\

    $ \mathcal{D}_{min SR_i} = min(\mathbf{D}_{SR});$  \\

    $\mathbf{D}_{min}=[ \mathbf{D}_{min}  \mathcal{D}_{min SR_i}];$ \\

$\mathbf{end}$ \\
\\

$\mathbf{for}$  i=1:N \\

    ~~ $\mathbf{D}_{RD}=[ ];$ \\

   ~~  $\mathbf{for}$  l= 1: $2^M - 1$ \\

      ~~~~  $\mathbf{for}$  n= l+1: $2^M$ \\

          ~~~~   $\mathcal{D}_{R_iD}=  \norm{\sqrt{ E_s/M} \mathbf{H}_{r_i,d}\mathbf{x}_l - \sqrt{ E_s/M} \mathbf{H}_{r_i,d}\mathbf{x}_n}^2;$ \\

           ~~~~ $\mathbf{D}_{RD}=[ \mathbf{D}_{ RD}  \mathcal{D}_{R_iD}];$ \\

~~~~        $\mathbf{end}$\\

    ~~$\mathbf{end}$\\

      ~~$ \mathcal{D}_{min R_iD} = min(\mathbf{D}_{RD});$
     \\

    ~~$\mathbf{D}_{min}=[ \mathbf{D}_{min}  \mathcal{D}_{min R_iD}];$ \\

$\mathbf{end}$ \\
\\

[$\mathbf{distance}$,$\mathbf{indice}$] $=$ sort( $\mathbf{D}_{min}$) ;\\
\\

$\mathbf{if}$ $\mathbf{indice}$$(2 \times N)$ $\leq$  N \\

 ~~  k= $\mathbf{indice}$$(2 \times N)$; \\

  ~~ $relay_{selected}$  $=$ $ R_k$; \\

 ~~   $\mathbf{ else}$  \\

  ~~~~ k = $\mathbf{ indice}$$(2 \times N)$ - N; \\

    ~~~~    $relay_{selected}$ $=$ $R_{k}$ ;\\

$\mathbf{end}$ \\

\\

 $\mathcal{D} _{max min SR-RD}= max(\mathbf{distance});$
\\
   \\
$\mathbf{D}_{SD}=[ ];$ \\

$\mathbf{for}$  l= 1: $2^M - 1$ \\

 ~~   $\mathbf{for}$  n= l+1: $2^M$ \\

  ~~~~      $\mathcal{D}_{SD}=  \norm{\sqrt{ 2 \times E_s/M} \mathbf{H}_{s,d}\mathbf{x}_l - \sqrt{ 2 \times E_s/M} \mathbf{H}_{s,d}\mathbf{x}_n}^2;$
      \\

 ~~~~       $\mathbf{D}_{SD}=[ \mathbf{D}_{SD}  \mathcal{D}_{SD}];$  \\

~~    $\mathbf{end}$ \\

$\mathbf{end}$  \\
\\

$ \mathcal{D}_{min SD} = min(\mathbf{D}_{SD});$
      \\
\\

$\mathbf{if}$  $\mathcal{D}_{min SD} \geq \mathcal{D}_{max min SR-RD}$  \\

~~    mode$=$ "Direct transmission mode"; \\

  ~~  $\mathbf{ else}$  \\

~~~~mode$=$ "Max-Link mode" ;\\

$\mathbf{end}$

\end{tabular}
\end{table}

\subsection{Calculation of relay selection metric}

In the first step we calculate the metric $\mathcal{D}_{SR_i}$
related to the $SR$ channels of each relay $R_i$ in Max-Link mode:
\begin{eqnarray}
  \mathcal{D}_{SR_i}=  \norm{\sqrt{ E_s/M} \mathbf{H}_{s,r_i}\mathbf{x}_l - \sqrt{ E_s/M} \mathbf{H}_{s,r_i}\mathbf{x}_n}^2,
  \label{eq:7}
\end{eqnarray}
where "l" is different from "n",  $\mathbf{x}_l$  and $\mathbf{x}_n$
represent each possibility of the vector formed by $M$ symbols. As
an example, if $M = 2$, $\mathbf{x}_n$  and $\mathbf{x}_l$ may be [0
0]$^T$, [0 1]$^T$, [1 0]$^T$ or [1 1]$^T$.

This metric is calculated for each one of the $C_2^{2^M}$
(combination of $2^M$  in $2$) possibilities. So, in the example (if
$M = 2$), we have $C_2^4= 6$ possibilities. After calculating this
metric for each of the possibilities, we store the information
related to the smallest metric ($\mathcal{D}_{min SR_i}$), for being
critical (a bottleneck) in terms of performance, and thus each relay
will have a minimum distance associated with its SR channels.

In the second step we calculate the metric $\mathcal{D}_{R_iD}$
related to the $RD$ channels of each relay $R_i$:
\begin{eqnarray}
    \mathcal{D}_{R_iD}=  \norm{\sqrt{ E_s/M} \mathbf{H}_{r_i,d}\mathbf{x}_l - \sqrt{ E_s/M} \mathbf{H}_{r_i,d}\mathbf{x}_n}^2,
    \label{eq:8}
\end{eqnarray}
where "l" is different from "n". This metric is calculated for each
one of the $C_2^{2^M}$  possibilities. After calculating this metric
for each one of the possibilities, we store the information related
to the minimum distance ($\mathcal{D}_{min R_iD}$), and thus each
relay will have a minimum distance associated with its RD channels.

In the third step, after calculating the metrics $\mathcal{D}_{min
SR_i}$  and $\mathcal{D}_{min R_iD}$ for each of the relays, as
described previously, we look for the largest (maximum) value of the
minimum distance:
\begin{eqnarray}
    \mathcal{D} _{\max \min SR-RD}= \max(\mathcal{D}_{\min SR_i}, \mathcal{D}_{\min R_iD}),
    \label{eq:9}
    \end{eqnarray}
where "i" is the index of each relay (1,2,...,N). Therefore, we
select the relay that is associated with this $\mathcal{D} _{max min
SR-RD}$. This relay will be selected for reception or transmission,
depending on this metric  is associated with the SR or RD channels,
respectively. Note that most quantities are assumed known but for
parameter estimation techniques the reader is referred to
\cite{scharf,bar-ness,pados99,reed98,hua,goldstein,santos,qian,delamarespl07,xutsa,delamaretsp,kwak,xu&liu,delamareccm,wcccm,delamareelb,jidf,delamarecl,delamaresp,delamaretvt,jioel,delamarespl07,delamare_ccmmswf,jidf_echo,delamaretvt10,delamaretvt2011ST,delamare10,fa10,lei09,ccmavf,lei10,jio_ccm,ccmavf,stap_jio,zhaocheng,zhaocheng2}

\subsection{Calculation of the metric for direct transmission}

In this step we calculate the metric $\mathcal{D}_{SD}$ related to
the $SD$ channels for the direct transmission mode:
\begin{eqnarray}
    \mathcal{D}_{SD}=  \norm{\sqrt{ 2 E_s/M} \mathbf{H}_{s,d}\mathbf{x}_l - \sqrt{ 2 E_s/M} \mathbf{H}_{s,d}\mathbf{x}_n}^2,
    \label{eq:10}
\end{eqnarray}
where "l" is different from "n". This metric is calculated for each
of the $C_2^{2^M}$  possibilities. After calculating this metric for
each of the possibilities, we store the information related to the
minimum distance ($\mathcal{D}_{\min SD}$),  associated with SD
channels.

\subsection{Comparison of metrics and choice of transmission mode}

After calculating all the metrics associated to the SR and RD
channels, finding $\mathcal{D}_{max min SR-RD}$ and calculating the
metrics associated to the SD channels and finding $\mathcal{D}_{min
SD}$, we compare these parameters and select the transmission mode:

    - If $\mathcal{D}_{\min SD} \geq \mathcal{D}_{\max \min SR-RD}$, we select "Direct transmission mode",

    - Otherwise, we select "Max-Link mode".

If we do not consider the possibility of operating in "Direct
Transmission" mode (considering only the "Max-Link" mode), we have
the "Max-Link" scheme instead of the the proposed "Switched
Max-Link" scheme. Section IV illustrates and discusses the
simulation results of the  proposed "Switched Max-Link", the
"Max-Link" and the "conventional MIMO" (direct transmission)
schemes.

\section{Simulation Results}

The proposed Switched Max-Link scheme is considered in a network
with a source node, ten relays (N = 10) and one destination node and
$M$ antennas in each node, forming a number of $10\times M$ SR links
for reception and $10\times M$  RD  links for transmission. Each
relay has its buffer size equal to 20 packets. We have considered
the maximum minimum distance as the relay selection criterion. Since
each packet received by the relay is not necessarily transmitted to
the destination in the next time slot, it was necessary to insert in
the preamble of each packet the order information (its position in
the binary format, ranging from 1 to 10000).

We assume that the transmitted signals belong to a BPSK
constellation, that we have unit power channels ($\sigma_{ s,r}^2$
$=$ $\sigma_{ r,d}^2$ $=$  $\sigma_{ s,d}^2$ $= 1$) , $N_0 =1$ (AWGN
noise power) and $E_S = E_{r_j} = E$ (total energy transmitted per
time slot). The transmit signal-to-noise ratio SNR ($E/N_0$) ranges
from 0 to 16 dB and the performances of the transmission schemes
were tested for 10000 packets, each containing 100 symbols.

\begin{figure}[!h]
\centering
\includegraphics[scale=0.4]{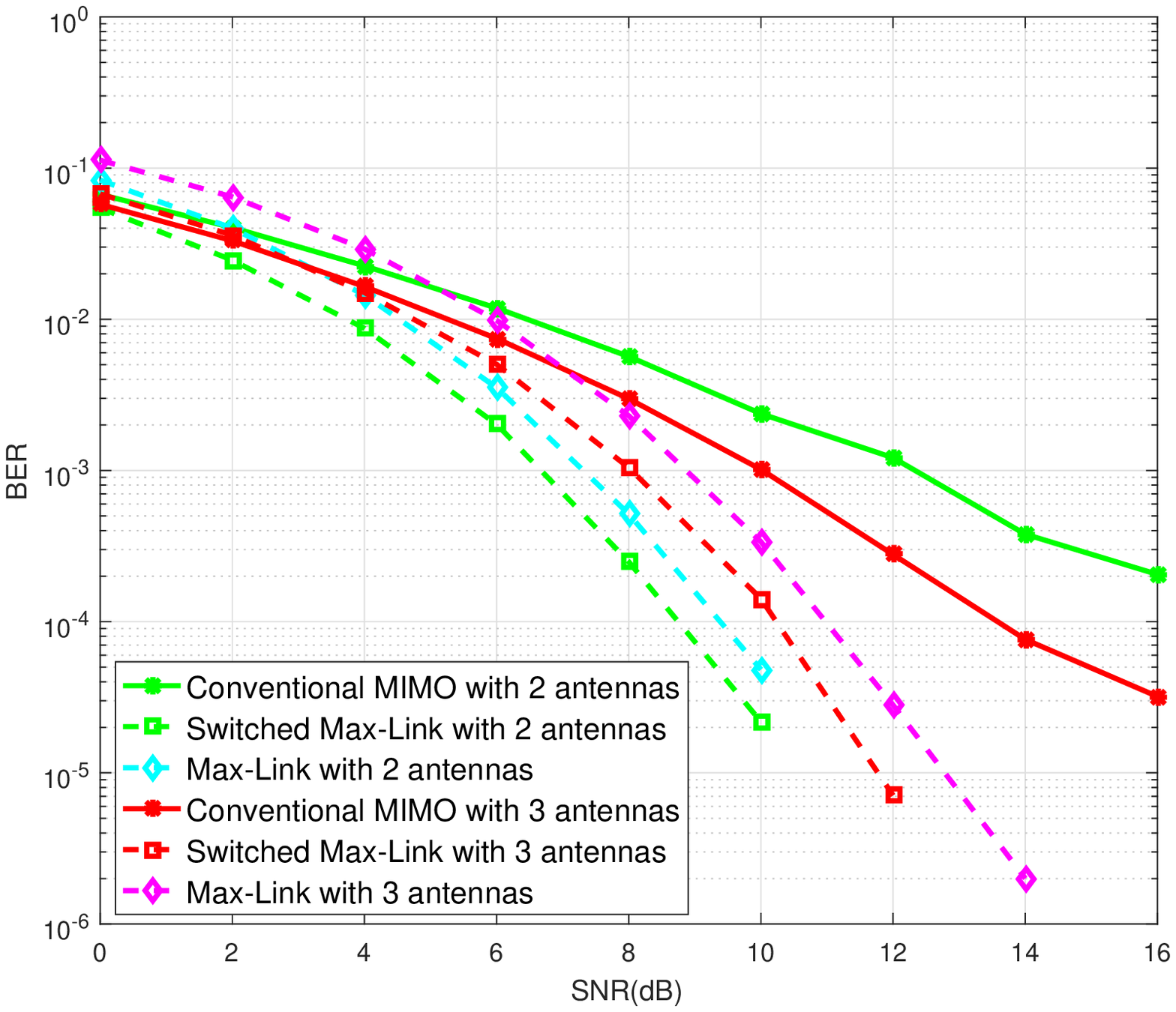}
\caption{Switched Max-Link, Max-Link and Conventional
multiple-antenna (direct transmission) BER  performance.}
\label{fig:berMaxlink}
\end{figure}

Fig. \ref{fig:berMaxlink} shows the Switched Max-Link, the Max-Link
and the conventional MIMO (direct transmission) BER  performance
comparison. By the analysis of this result,  it is observed that the
performance of the Max-Link scheme with 3 antennas  is worse than
the performance of the conventional MIMO scheme for a SNR less than
7 dB. Nevertheless, the performance of the proposed Switched
Max-Link scheme was quite  better than the performance of the
conventional direct transmission) for almost the total range of SNR
tested. It is observed, as expected, that the performance of the
proposed Switched Max-Link scheme was  better than the performance
of the the Max-Link scheme, with $M$ equal to  2 and 3. So we
decided to simulate the proposed scheme and compare it's performance
with the performance of the conventional MIMO for other values of
$M$.

\begin{figure}[!h]
\centering
\includegraphics[scale=0.4]{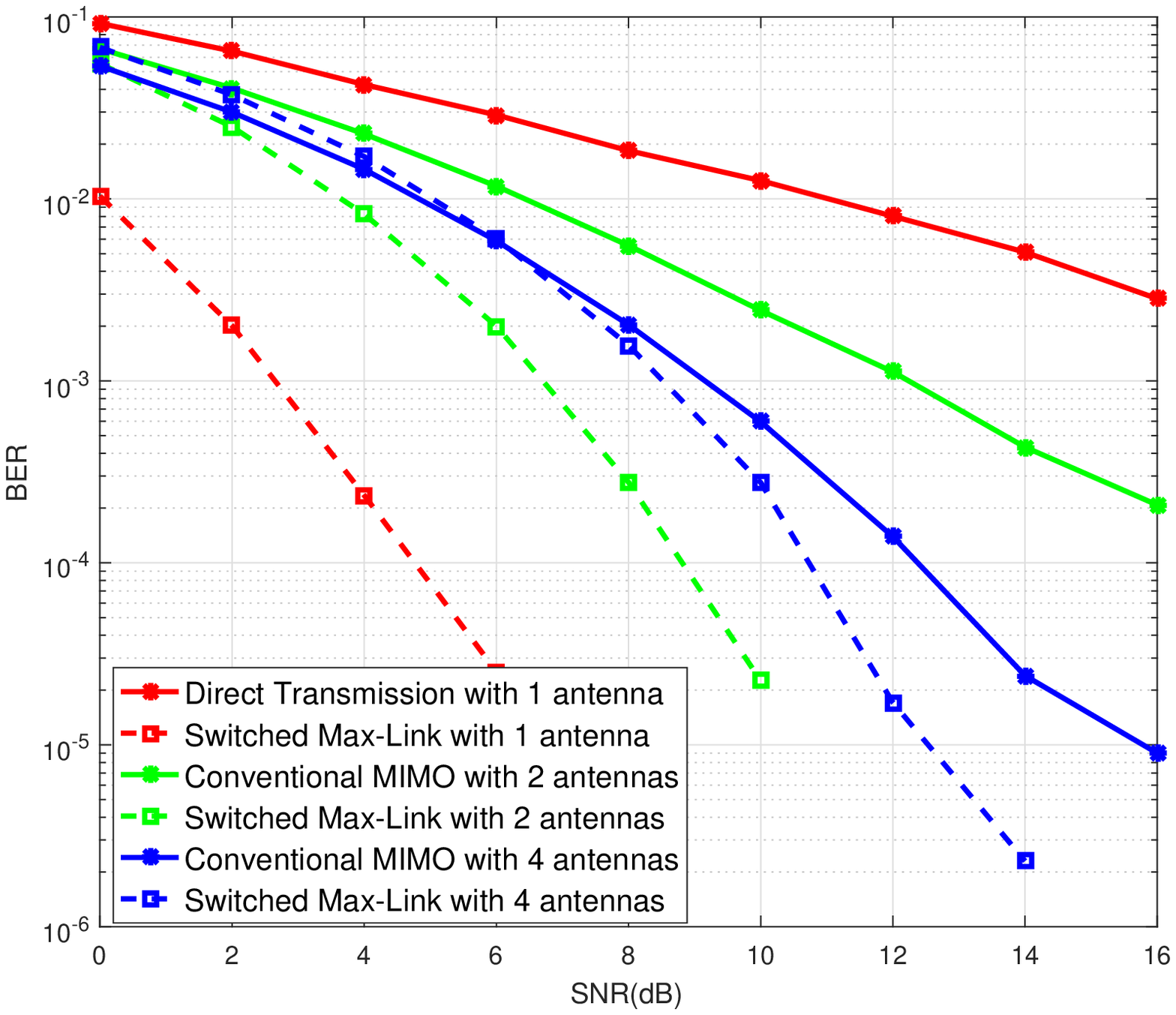}
\caption{Switched Max-Link and Conventional MIMO (direct transmission) BER  performance comparison}
\label{fig:ber}
\end{figure}

Fig. \ref{fig:ber} shows the Switched Max-Link and the conventional
direct transmission BER  performance comparison. By the analysis of
this result,  it is observed, as expected, that the performance of
the proposed Switched Max-Link scheme was quite better than the
performance of the conventional MIMO (direct transmission) to the
same number of antennas, with $M$ equal to 1, 2 and 4. It is also
observed that by increasing the number of antennas, a considerable
improvement in BER performance is obtained for the direct
transmission scheme, but it did not happen for the proposed scheme.

It is observed that by increasing the number of antennas in the
proposed scheme, the BER performance falls a bit. To illustrate
this, we will talk about what happens when we increase the number of
antennas from 1 to 2. The Switched Max-Link scheme, when operating
with only 1 antenna on each node, seeks and selects the best channel
for reception or transmission among the 20 channels (the best SR
channel among 10 of them or the best RD channel among 10 of them),
or select the SD channel if it worths. Otherwise, with two antennas,
this scheme seeks and selects the best pair of channels for
reception or transmission between 20 pairs (the best pair of SR
channels among 10 of them or the best pair of RD channels among 10
of them), or select the pair of SD channels if it worths. So it may
occur that even choosing the best pair of channels, the metric
(maximum minimum distance) obtained by operating with 2 antennas may
be not so good  as the metric obtained by operating with only 1
antenna. This has been checked experimentally.

\section{Conclusions}

In this paper we have presented the benefits of using \textit{
buffers} and multiple antennas for the design of half-duplex
decode-and-forward relaying protocols in cooperative communication
systems, introducing a new relay selection criterion called maximum
minimum distance based on the ML criterion. Moreover, a new
cooperative protocol using multiple antennas that combines switching
and Max-Link called Switched Max-Link has been proposed. The
performance of the proposed "Switched Max-Link" was evaluated
experimentally and outperformed the conventional direct transmission
and the existing Max-Link scheme.

\end{document}